\documentclass[aip,reprint]{revtex4-1}

\usepackage{graphicx}
\usepackage{dcolumn}
\usepackage{bm}
\usepackage[utf8]{inputenc}
\usepackage[T1]{fontenc}
\usepackage{mathptmx}
\usepackage{etoolbox}

\makeatletter
\def\@email#1#2{%
 \endgroup
 \patchcmd{\titleblock@produce}
  {\frontmatter@RRAPformat}
  {\frontmatter@RRAPformat{\produce@RRAP{*#1\href{mailto:#2}{#2}}}\frontmatter@RRAPformat}
  {}{}
}%
\makeatother

\draft 
\begin{document}
\title{Non-Line-of-Sight Passive Acoustic Localization Around Corners\\} 
\author{Jeremy Boger-Lombard}
\author{Yevgeny Slobodkin}
\author{Ori Katz}
 \email{Orik@mail.huji.ac.il.}
\affiliation{Department of Applied Physics, Hebrew University of Jerusalem 9190401, Israel.}

\date{\today}
\begin{abstract}
Non-line-of-sight (NLoS) imaging is an important challenge in many fields ranging from autonomous vehicles and smart cities to defense applications. Several recent works in optics and acoustics tackle the challenge of imaging targets hidden from view (e.g. placed around a corner) by measuring time-of-flight (ToF) information using active SONAR/LiDAR techniques, effectively mapping the Green functions (impulse responses) from several sources to an array of detectors. Here, leveraging passive correlations-based imaging techniques, we study the possibility of acoustic NLoS target localization around a corner without the use of controlled active sources. We demonstrate localization and tracking of a human subject hidden around the corner in a reverberating room, using Green functions retrieved from correlations of broadband noise in multiple detectors. Our results demonstrate that the controlled active sources can be replaced by passive detectors as long as a sufficiently broadband noise is present in the scene. 
\end{abstract}
\maketitle 

Non-line-of-sight (NLoS) imaging techniques have important applications in the fields of autonomous vehicles navigation and remote sensing\cite{faccio2020non}. NLoS techniques aim to localize, track, and image targets hidden from view by recording 'multiply-bounced' reflected waves, i.e. waves that reflect off a directly visible surface, such as a wall, towards the hidden target, and back from it to a detector array by another reflection. 
In the last decade there have been great advancements in the field, enabling high resolution NLoS imaging and tracking in real-time for a variety of applications using both light and sound \cite{faccio2020non,velten2012recovering,o2018confocal,liu2019non,lindell2019wave,boger2019passive,nam2021low,lindell2019acoustic}. 

In the optical domain, time-of-flight (ToF) techniques, achieve centimeter-scale lateral resolution by computational back-projection reconstruction \cite{o2018confocal,liu2019non,lindell2019wave,boger2019passive,nam2021low}. However, since in the optical domain the reflections from most common surfaces are diffuse reflections, due to the surface roughness being large compared to the optical wavelength,
the quartic falloff of the multi-bounce diffuse reflections fundamentally limit the imaging range. In addition, many real life applications, such as in automotive and indoor tracking of subjects, do not require the centimeter-scale resolution achievable via optical NLoS techniques, making acoustic-based NLoS techniques attractive.

When acoustic waves are considered \cite{lindell2019acoustic}, the optically-rough surfaces of e.g. white-painted walls, become effectively flat reflective mirrors due to the considerably longer acoustic wavelength ($\lambda\approx1m-10cm$ for acoustic frequencies of 300Hz-3KHz). The specular reflections of audible-frequency waves from most ordinary walls can then straightforwardly reveal the mirror image of the hidden targets by conventional beam-forming back-projection techniques  \cite{lindell2019acoustic}, similar to the ones used in ultrasound echography. Furthermore, in the acoustic domain, the direct measurement of the acoustic fields is performed using conventional off-the-shelf microphones, and does not require specialized ultrafast detectors or interferometric techniques, as used in the optical domain.

Acoustic NLoS localization of active sources, such as speakers, has been long demonstrated using either reflected waves\cite{mak2009non, kitic2014hearing}, or waves refracted by a cornered edge of an occluder \cite{singh2012non}. Recently, Lindell et al. have demonstrated NLoS localization and imaging of passive reflectors in an anechoic chamber by applying a multi-bounce ToF approach, utilizing an array of microphones and speakers emitting strong chirped pulses \cite{lindell2019acoustic}. Specifically, the pulsed emissions from each of the speakers and consecutive measurements of the reflected waves by the microphones array have allowed the retrieval of a set of speaker-microphone Green functions. These were then used to reconstruct the hidden scene by beam forming back-projection. 

Here, we study the possibility of retrieving the same set of temporal Green functions \textit{passively}, i.e. without emitting controlled acoustic waveforms. To achieve this, we leverage the ideas of passive imaging \cite{snieder2010imaging,snieder2006theory,duvall1993time,weaver2001ultrasonics,lobkis2001emergence,shapiro2005high,davy2013green,badon2015retrieving,garnier2016passive} to estimate the Green functions from cross-correlations of ambient broadband noise, using only an array of microphones. 
We demonstrate localization of a human subject around the corner in a reverberating concrete-walled room containing several uncontrolled broadband noise sources. In our experiments, random diffuse signals reveal pulse-echo like reflected signals via temporal cross-correlations between pairs of microphones in the array, which are then used as the estimates of the Green functions to faithfully estimate the hidden targets positions.

\begin{figure*}[hbt!]
\includegraphics{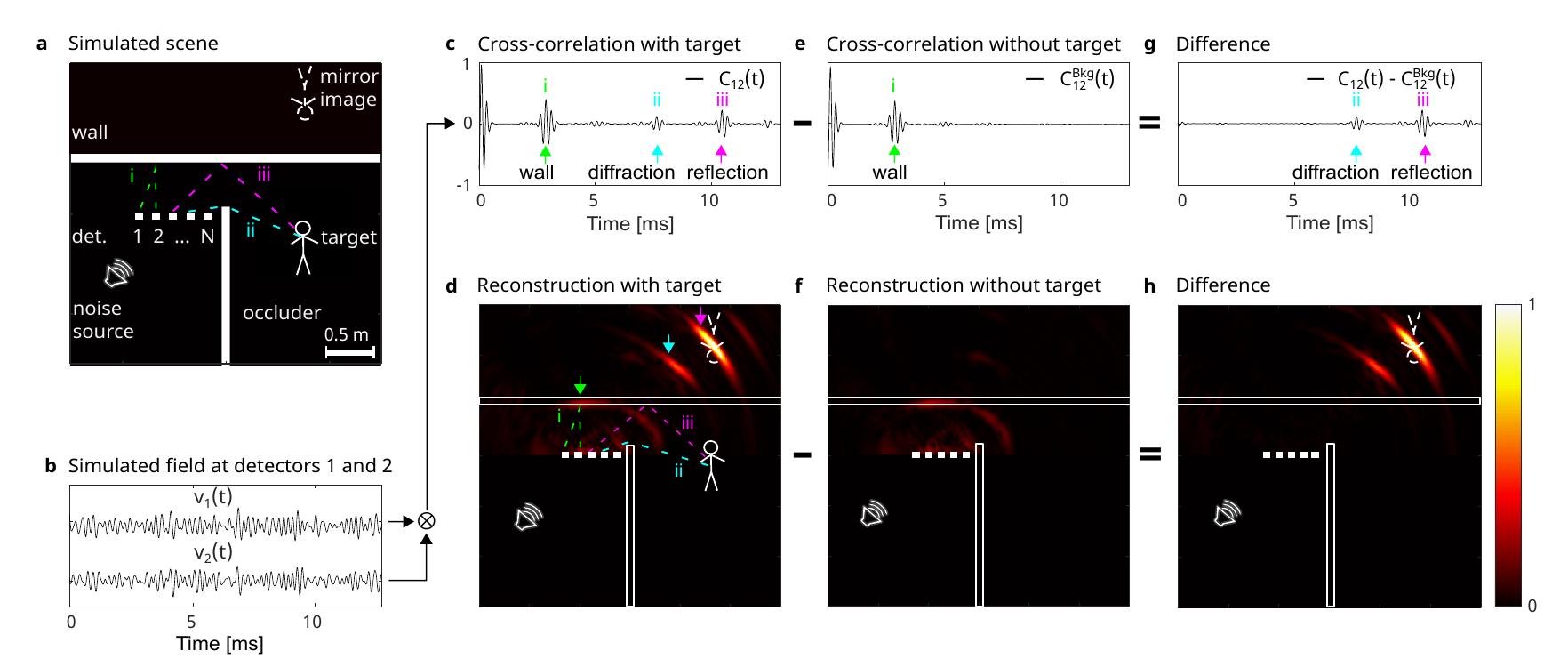}
\caption{\label{fig:1} Passive NLoS localization process (numerical example). \textbf{a} The simulated scene (top view): a target is hidden behind an occluder. A 16-detectors array records the continuous broadband noise emitted by a nearby uncontrolled source, which reverberates in the scene. The recorded noise contains directly arriving signals, single reflections (in green,i), diffracted reflections  (cyan, ii), and multiple reflections (magenta, iii) allowing NLoS localization. \textbf{b} Noise fields $v_1(t), v_2(t)$ recorded by detectors 1,2, respectively. \textbf{c} Cross-correlation of the recorded fields $C_{12}(t)$, reveals pulse-echo like ToF information containing: (i) direct reflections from the wall; (ii) fields that originate from diffraction by the occluder edge to the target; (iii) fields that reflect by the wall to the target and back. These are used for direct localization of the target mirror image. \textbf{d} Delay-and-sum beam-forming reconstruction from 16x16 cross-correlations (as in (\textbf{c})) for all detector pairs. The positions of the wall (green arrow), the target mirror image (magenta arrow), as well as the edge diffraction artifact (cyan arrow) are visible. \textbf{e,f} Same as (c,d), for a scene without the hidden target. \textbf{g} Difference between the cross-correlations of (\textbf{c}) and (\textbf{e}). \textbf{h} Difference between (\textbf{d}) and (\textbf{f}) shows only the hidden target contributions.}
\end{figure*}

Our work is based on passive correlation imaging, also known as coda-interferometry\cite{snieder2006theory}. The working principle of coda-interferometry is that by correlating ambient noise one can reproduce the Green function which contains the same ToF information measured in active pulse-echo experiments. The idea was first put to use in helioseismology for extracting the travel time of acoustic waves from temporal cross-correlations of the intensity fluctuations on the solar surface \cite{duvall1993time}. Lobkis and Weaver have shown that the autocorrelation function of ultrasound noise measurements reveals the same waveform as the one measured in a single transducer pulse-echo experiment\cite{weaver2001ultrasonics}, and that the cross-correlation between two registrations of the diffuse noise field at two arbitrary points in space can reveal the Green's function between these points\cite{lobkis2001emergence}. The approach was also put to use in geophysics\cite{shapiro2005high}, microwave\cite{davy2013green}, and in optical studies of complex media\cite{badon2015retrieving}. 

As passive correlation allows to acquire the same ToF information as obtained in active pulse-echo experiments, it could be used, in principle, to localize hidden targets in a NLoS scenario in the same fashion as conventional ToF measurements \cite{velten2012recovering, lindell2019acoustic}. Thus, one can utilize uncontrolled broadband noise sources for passive NLoS imaging of reflective targets, in a similar fashion to the use in direct passive imaging \cite{garnier2016passive}. This is the goal we were set to demonstrate in this work.

\begin{figure*}[hbt!]
\includegraphics{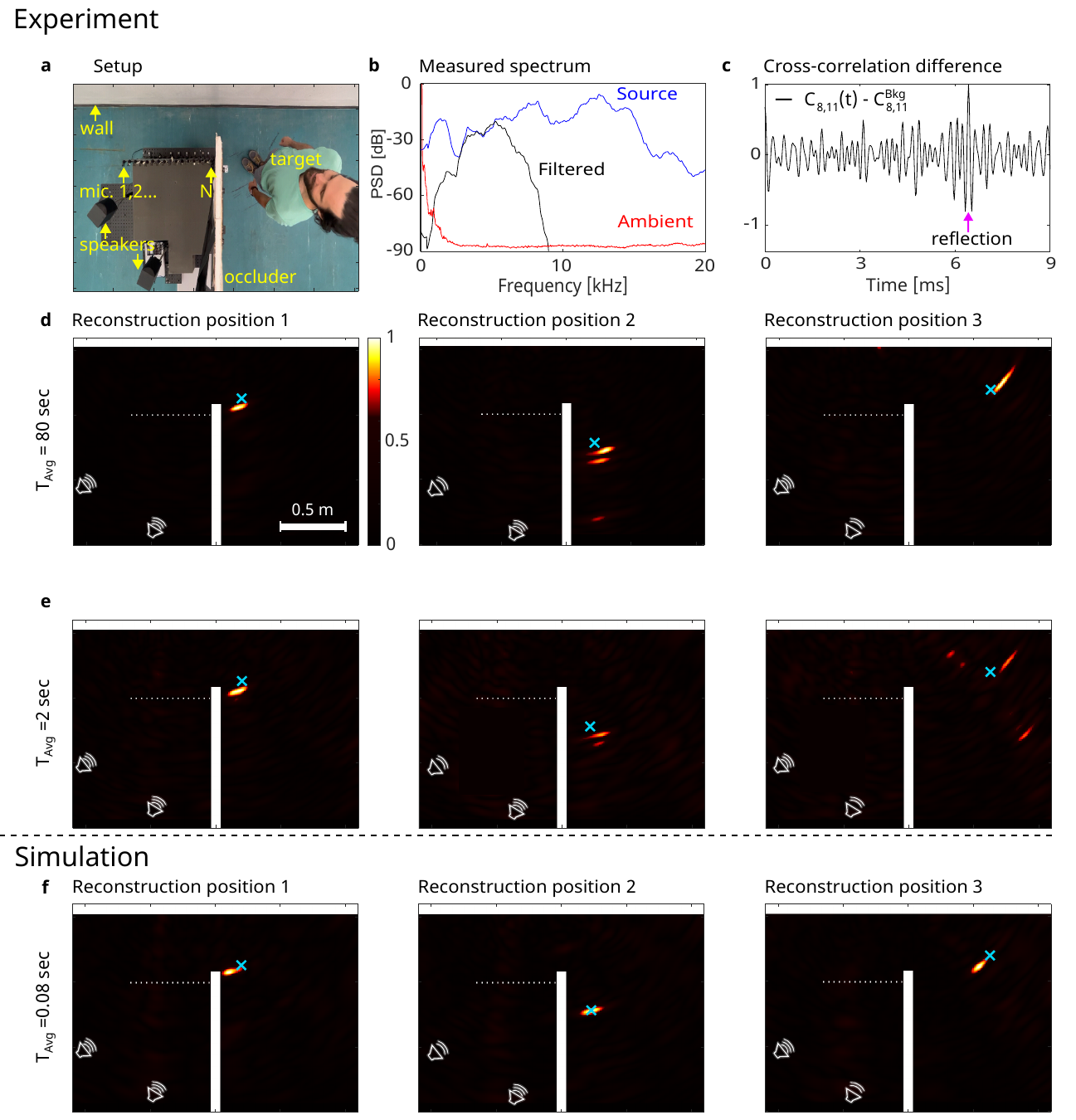}
\caption{\label{fig:2} Experimental passive acoustic NLoS localization and tracking of a hidden subject around-the-corner. \textbf{a} Setup (top view): A subject hides behind an occluder. Two uncorrelated speakers emit broadband random noise. A linear array of $N=16$ microphones records the acoustic pressure fields. \textbf{b} Power spectral density (PSD) of the raw measured signal in microphone number 1 (blue curve), the bandpass-filtered signal used for reconstructions (black curve), and the ambient noise when the sources are off (red curve). \textbf{c} Difference in cross-correlations of a single pair of microphones when the target is present and when the target is absent. Arrow marks the desired double-reflection (wall-target-wall) that provides the target position. \textbf{d,e} Experimental results: beamforming reconstructions from experimental cross-correlations locating a person at 3 different positions around the corner. Integration times: $T_{Avg} = 80 sec$ (\textbf{d}), and $T_{Avg} = 2 sec$ (\textbf{e}). A cyan cross marks the true positions. The reconstructions are mirrored with respect to the wall. \textbf{f} Numerical results of a simulated scenes without reverberations or measurement noise, $T_{avg} = 0.08 sec$.}
\end{figure*}

The principle of our approach and the setup for realizing it are depicted in Fig.~\ref{fig:1}a, accompanied by a numerically simulated sample result (Fig.~\ref{fig:1}b-h, see Supplementary Material). We consider a simplified scenario, where a hidden target is outside the line of sight for both a microphone array and a broadband uncontrolled noise source (Fig.~\ref{fig:1}a). A broadband acoustic noise field emitted by the noise source is reflected off the target either by reflection from the relay wall (iii, depicted by a magenta dashed line in Fig.~\ref{fig:1}a) or by diffraction from the occluding wall edge (ii, depicted in cyan in Fig.~\ref{fig:1}a). A detector array composed of $N$ microphones records these reflected fields, in addition to reflections from the walls in the scene (e.g. (i) depicted in green), and the direct arriving waves from the noise source. 
The waveforms $v_j(t)$ $j=1..N$, recorded at the different detectors are given in Fig.~\ref{fig:1}b. While seemingly random, the cross-correlation, $C_{ij}(\tau )$ between each pair $i,j$ of the recorded waveforms reveals pulse-echo like ToF information (Fig.~\ref{fig:1}c): 

\begin{equation}\label{eq1}
C_{ij}(\tau ) = \frac{1}{T_{avg}} \int_{0}^{T_{avg}} v_i(t)v_j(t+\tau)dt     
\end{equation}

\begin{figure*}
\includegraphics[width=\textwidth,height=\textheight,keepaspectratio]{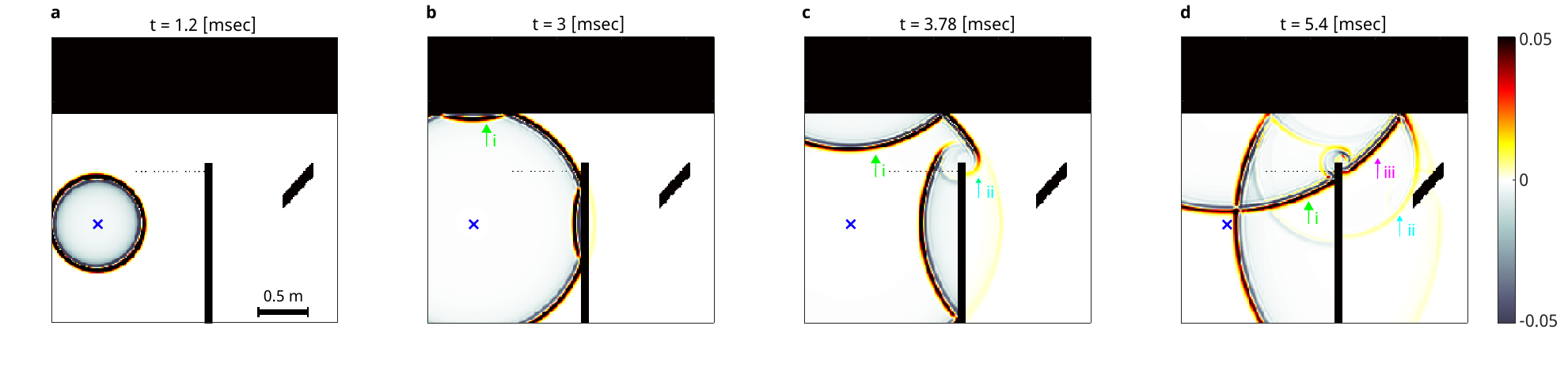}
\caption{\label{fig:3} Numerical study of the wave propagation in the considered scene reveal the various contributions in the measured signals.
\textbf{a-d} Acoustic pressure distribution of the propagating wave from a short pulsed source (blue x), at four different propagation times.
\textbf{a} Free-space spherical wave propagation before reaching any reflectors/occluders.
\textbf{b} First reflections from the wall (green arrow) and occluder.
\textbf{c} At later time, the reflection from the wall (green arrow, i) propagates towards the target. Diffraction of the direct wave from the occluder edge generates a weak diffracted wave propagating towards the target (cyan arrow, ii). 
\textbf{d} The edge-diffracted wave hits the target (cyan arrow, ii).
The wavefront reflected from the wall arrives both directly at the detector array (green arrow, i), and at a later time to the target (magenta arrow, iii).}
\end{figure*}

Where $T_{avg}$ is the recording (averaging) time, and $\tau$ is the variable computed lag time between the two waveforms. This simple post-processing provides an estimate of the Green function between the two detectors. The longer is $T_{avg}$ the better is the estimate \cite{seats2012improved}. Since the cross-correlated data is approximately equivalent to a measurement of a pulsed source and detector pair \cite{lobkis2001emergence}, it can be beam-formed back to form an image by conventional delay and sum beamforming \cite{friis1937multiple,perrot2021so} (Fig.~\ref{fig:1}d), assuming that the reflecting 'relay wall' is a flat mirror, which is a good approximation for most common indoor walls.
The presence of the multiple reflections that do not originate from the target result in strong reconstructed features that are not related to the target (Fig.~\ref{fig:1}d), but originate from the static walls in the scene. These contributions can be subtracted using an additional identical measurement performed without the target present in the scene (Fig.~\ref{fig:1}e,f), where only the contributions of the walls are present (a background measurement). Taking the difference between the cross-correlation of the measurements with and without a target leaves only the target-related signals (Fig.~\ref{fig:1}g). Beam-forming using these signals allows localizing the position of the target mirror-image (Fig.~\ref{fig:1}h). A reconstruction artefact originating from early-arriving signals appear in the beam-formed image (marked by a cyan arrow in Fig.~\ref{fig:1}h). This artefact originates from signals that diffract off the cornered edge of the barrier rather than the relay wall in either the detection or sonification paths (Fig.~\ref{fig:1}g (ii, cyan arrow)). A more detailed analysis of this diffraction artefact is given below (Fig.~\ref{fig:3}).

Figure~\ref{fig:2} presents experimental results of passive acoustic localization around the corner. A photo of the experimental setup is given in Fig.~\ref{fig:2}a: A human subject is hidden around the corner from a linear array of $N=16$ microphones that records the acoustic fields from two uncontrolled broadband sources (Fig.~\ref{fig:2}c).
The broadband spectrum of the raw measured signal of a single microphone is given in Fig.~\ref{fig:2}b. We calculate the pair-wise cross-correlations between the measured signals after band-pass filtering the raw recorded signals with a Gaussian filter of central frequency $f_0 = 5.3kHz$ and a full width at half max (FWHM) bandwidth of $\Delta f_{FWHM} = 1.8kHz$. 
Repeating the cross-correlations calculation for signals acquired with and without the subject present, and taking their difference reveals a pulse-echo like ToF information with a peak at the expected delay time (Fig.~\ref{fig:2}c). Applying delay-and-sum beamforming on the $N^2$ cross-correlations traces, and flipping the reconstructed (mirror) image vertically in respect to the relay wall, localizes faithfully the subject's position in several locations by analyzing different $80s$-long temporal segments of a single recording (Fig.~\ref{fig:2}d, true positions marked by blue crosses). Using shorter recorded segments of $T_{avg} = 2sec$ still reveals the correct positions of the hidden target, with more artefacts present (Fig.~\ref{fig:2}e). Numerical simulation of the simplified experimental scene without  the presence of noise and additional reflections that are outside the shown field of view, shows good qualitative agreement with the experimental reconstructions (Fig.~\ref{fig:2}f).

To provide a more in depth analysis and understanding of the origins of the diffraction artefact present in Fig.~\ref{fig:1}(g,h), we display in Fig.~\ref{fig:3} four snapshots of a simulated propagated impulse field from one noise source. The simulated results have been obtained by a two-dimensional FDTD simulation (k-Wave \cite{treeby2018rapid}, see supplementary material): In Fig.~\ref{fig:3}a, the free-space propagation results in a perfect spherical wavefront. When the pulse front hits the walls (Fig.~\ref{fig:3}b) it is reflected from the relay-wall (green arrow, i) and the occluding barrier. 
Shortly after (Fig.~\ref{fig:3}c) two phenomena can be observed: The first is the propagation of the reflected wave from the relay wall (green arrow,i), and the second is the weak, but non-negligible, 'knife-edge' diffraction from the edge of the occluding barrier (cyan arrow, ii). 
Finally, at later times (Fig.~\ref{fig:3}d), while the wave reflected from the relay wall continues to propagate towards the target (magenta arrow, iii), the weak knife-edge diffracted wave already arrives to the target (cyan arrow). The contribution from both of these signals will be eventually recorded by the detectors. While the diffracted peak arrives at an earlier time (cyan arrow in Fig.~\ref{fig:1}c,g) than the signal reflected from the relay wall (magenta arrow in Fig.~\ref{fig:1}c,g), only the latter will yield the correct position of the target when conventional beam-forming is used for reconstruction. Nonetheless, knowledge of the visible scene geometry can be used to take into account the contribution of such knife-edge diffraction signals to improve the reconstruction. Removing undesired artifacts and improving the SNR in the reconstructed image, can be achieved by diffraction and reflections aware localization \cite{an2019diffraction}. 

To summarize, we have demonstrated an approach that allows to passively localize and track a person hidden around a corner using conventional off-the-shelf microphones and uncontrolled broadband sources. The spatial localization accuracy is dictated by the ToF temporal resolution, which is given by the temporal width of the cross-correlation peak. For a broadband source, this width is given by the source coherence time $t_c \approx 1/\Delta f$, where $\Delta f$ is the source spectral bandwidth. Each single ToF measurement from temporal cross-correlation between two detectors, localizes the target on an ellipsoid surface (or a sphere in the case of the autocorrelation of a single detector) with an axial resolution of $dr \approx c_s/2\Delta f$. Where $c_s$ is the speed of sound. Assuming a perfect retrieval of the Green functions, the final reconstruction resolution is the same as for active SONAR experiments \cite{lindell2019acoustic}. In practice, the finite recording time will result in noisy cross-correlations and thus to reconstruction clutter artefacts (Fig.~\ref{fig:2}).

Our method is based on Green function retrieval from temporal cross-correlations of broadband noise. In most works the noise field is assumed to be diffuse and isotropic\cite{weaver2001ultrasonics}, which may be indeed the case for strongly reverberant rooms. In the case of an anisotropic noise field, e.g. where the waves traveling in the medium are arriving mainly from one-sided half plane, the Green function retrieval would result in a one sided projection of either $G(x_i,x_j,t)$, or $ G(x_i,x_j,-t)$ \cite{lin2009eikonal}. In our experiments the field is not entirely diffuse, and we have noticed differences in the reconstructions depending on the exact placement of the non-isotropic noise sources.

The two main challenges in making the presented approach useful in practical scenarios are the relatively narrow bandwidth of common ambient noise (Fig.~\ref{fig:2}b, red curve), which result in a lower reconstruction resolution, and the current requirement for a relatively long averaging time. The averaging time can be lowered by using a larger number of detectors, and adapting advanced reconstruction approaches. Development of more advanced reconstruction algorithms that take into account the contributions of diffracted waves using the (known or measured) room geometry is expected to significantly improve the reconstruction fidelity. Similar data-driven approaches using neural networks have been recently put forward for optical NLoS reconstruction \cite{tancik2018data,chen2019steady}, and NLoS classification of individuals. \cite{caramazza2018neural}.

This work has received funding from the European Research Council under the European Union’s Horizon 2020 Research and Innovation Program grant number 101002406, and the Israel Science Foundation (grant number 1361/18).
\section*{AUTHOR DECLARATIONS}
\subsection*{Conflict of Interest}
The authors have no conflicts to disclose.
\subsection*{Author Contributions}
O.K. conceived the project. J.B.L. and O.K. designed the experimental setup. J.B.L. performed measurements and data analysis under the supervision of O.K. J.B.L. and Y.S. performed numerical simulations under the supervision of O.K. J.B.L. and O.K. wrote the manuscript.
\section*{DATA AVAILABILITY}
The data which support the findings of this study are available from the corresponding author upon reasonable request.

\section*{REFERENCES}

\bibliography{aiptemplate}

\clearpage

\section*{\textbf{Supplementary Material\\}}

\textbf{Experimental setup} 
The experimental setup is presented in Fig.~2a. The occluder was realized by a pair of acoustic drywall plates with two layers of Suprema - Tecsound pallet sandwiched between them. This 3 cm thick occluder was placed perpendicularly to the wall at a distance of 45 cm. Noise was generated by playing two different Gaussian random white noises through two audio speakers (MIYAKO Ltd, SL-800).
The microphone array consisted of 16 condenser microphones (BOYA, BY-M1) placed at a spacing of 4cm, and were sampled simultaneously at 40 kHz with 16 bit depth using a multichannel DAQ device (National Instruments, PXIe-6363). The array was placed at a distance of 53 cm from the wall, in parallel to it, the rightmost microphone was at a distance of 5 cm from the occluder.
A human subject served as the target in all experiments. 

\textbf{Numerical simulations} 
Simulations were performed using 'k-Wave', a 2D Finite-Difference Time-Domain (FDTD) simulation toolbox\cite{treeby2018rapid}. The simulations computed the propagation of a delta-like impulse pressure wave from each of the noise sources through the simulated scene to each of the microphones (Fig.~1a), yielding the Green functions from each source to each microphone.
The full simulated scene was represented by $400\times400$ pixels, with a pixel size of $\mathrm{1\;cm^2}$ representing a plane of $\mathrm{4 \;m\times 4\;m}$.
Free-space propagation through air was represented by a speed-of-sound of $\mathrm{345\;m/s}$ and density of $\mathrm{1.225\;kg/m^3}$.
The wall and occluder were represented by a $1.47m$ and $3 cm$ thick simulated regions having a density of $\mathrm{24.5\;kg/m^3}$, and speed of sound of $\mathrm{1500\;m/s}$, which yielded a high value of reflection coefficient and low transmission. 
The random noise sources were simulated by convolving the Green functions related to each source with a single random signal with a length of $7501\times10^3$ samples. The two random signals obtained for each microphone (from each of the two noise sources) were then summed, cropped to a finite measurement time, and were considered as the signal measured by this microphone. These 'measured' signals were then processed in the same manner as the measured experimental signals (Fig.~2b).
\end{document}